\documentclass[12pt]{article}
\usepackage[utf8]{inputenc}
\usepackage{booktabs,multirow}
\usepackage{amssymb, amsmath, mathbbol,amsfonts,dsfont,slashed,mathtools,array}
\usepackage{color}
\usepackage{graphicx}
\usepackage{float}
\graphicspath{{plots/}}
\usepackage[centerlast,footnotesize]{caption2}

\newcommand{\One}{1\kern-4.5pt1}

\newcommand{\be}{\begin{equation}}
\newcommand{\ee}{\end{equation}}

\begin{document}
\begin{center}
\begin{flushright}
MS-TP-15-03
\end{flushright}
\begin{flushright}
HIP-2014-33/TH
\end{flushright}
\begin{flushright}
January 2015
\end{flushright}
\vskip 10mm
{\LARGE
Hadron Wavefunctions as a Probe of a \\ Two Color 
Baryonic Medium\\
}
\vskip 0.3 cm
{\bf Alessandro Amato$^{a,b}$, Pietro Giudice$^c$ and Simon Hands$^a$}
\vskip 0.3 cm
$^a${\em Department of Physics, College of Science, Swansea University,\\
Singleton Park, Swansea SA2 8PP, United Kingdom.}
\vskip 0.3 cm
$^b${\em Department of Physics and Helsinki Institute of Physics, \\ P.O.Box 64, FI-00014 University of Helsinki, Finland.}
\vskip 0.3 cm
$^c${\em Institut f\"ur Theoretische Physik, Universit\"at M\"unster, M\"unster, Germany.}
\vskip 0.3 cm
\end{center}

\noindent
{\bf Abstract:} 
The properties of the ground state of two-color QCD at non-zero baryon chemical
potential $\mu$ present an interesting problem in strongly-interacting gauge
theory; in particular the nature of the physically-relevant degrees of freedom
in the superfluid phase in the post-onset regime $\mu>m_\pi/2$ still needs clarification. In
this study we present evidence for in-medium effects at high $\mu$ by studying
the wavefunctions of mesonic and diquark states using orthodox lattice
simulation techniques, made possible by the absence of a Sign Problem for the
model with $N_f=2$. Our results show that beyond onset the spatial extent of
hadrons decreases as $\mu$ grows, and that the wavefunction profiles are consistent with the existence of a
dynamically-gapped Fermi surface in this regime.  \vspace{1cm}

                                                                                
\vspace{0.5cm}
\noindent
Keywords: 
lattice simulation, chemical potential

\newpage
\section{Introduction}
QC$_2$D, the QCD-like gauge theory with gauge group SU(2), offers the simplest
opportunity for the controlled non-perturbative study of non-zero baryon charge
density via Monte Carlo lattice simulation, unhindered by a Sign Problem. The
reality of the quark determinant even with quark chemical potential $\mu\not=0$
(and the consequent positivity of the path integral measure for number of
flavors $N_f$ even) arises from the pseudoreality of the fundamental
representation of SU(2)~\cite{Hands:2000ei}.

Several studies, both analytical and numerical, have helped elucidate the phase
structure and highlighted the differences with the physical case of SU(3). At
low temperature $T\approx0$, the ground state density $n_q$ of quarks rises from
zero at a second-order {\sl onset\/} transition at $\mu=\mu_o=\frac{1}{2}m_\pi$,
permitting the existence of arbitrarily dilute baryonic matter. For current
quark masses sufficiently light to permit a separation of scales between $m_\pi$
and the next lightest hadron (either a $1^-$ $\rho$ meson or $1^+$ diquark
baryon), the regime $\mu\gtrsim\mu_o$ is described by chiral perturbation theory
($\chi$PT)~\cite{Kogut:2000ek}, and consists of a dilute gas of weakly-repelling
Bose diquarks, which condense to form a superfluid ground state via the
spontaneous breaking of the global U(1)$_B$ of baryon number. This regime has
been identified in simulations with staggered lattice
fermions~\cite{Hands:2000ei,Aloisio:2000if}. Higher densities have been probed
in simulations with Wilson lattice fermions further from the chiral
limit~\cite{Hands:2006ve,Cotter:2012mb}. Over a wide range of $\mu>\mu_o$
thermodynamic quantities such as $n_q$, pressure $p$ and energy density
$\varepsilon$~\cite{Cotter:2013oaa} are all found to scale approximately with
$\mu$ as expected of a degenerate system of quarks, ie. one in which the
available phase space is uniformly populated up to some characteristic momentum
scale $k_F\simeq\mu$, so that, eg.  \begin{equation} n_q=2N_fN_c\int_{\vert\vec
k\vert\leq k_F}{{d^3\vec k}\over{(2\pi)^3}}\simeq{{2N_f}\over{3\pi^2}}\mu^3.
\label{eq:SB} \end{equation} in complete contrast to the behaviour predicted by
$\chi$PT~\cite{Kogut:2000ek}.  However, the system remains in the same confined
state observed at $\mu=0$, as indicated by the near-vanishing of the Polyakov
line expectation $\langle L\rangle$. This state has some similarities with a
confined chirally-symmetric ground state originally predicted in the context of
large-$N_c$ QCD~\cite{McLerran:2007qj}; for this reason we have called it a {\sl
quarkyonic\/} regime. At the lowest temperatures studied, only for $\mu a\sim
O(1)$ has color deconfinement signalled by $\langle L\rangle>0$ been observed;
while the quark density $n_q$ is still much smaller than its saturation value
$2N_fN_c/a^3$ at this point, without simulations on a finer lattice we cannot
yet be confident this high density deconfined phase persists in the continuum
limit. Finally we note that the simulations suggest that the ground state is
superfluid, as signalled by a non-vanishing diquark condensate, for all
$\mu>\mu_o$; however in the quarkyonic regime the scaling of the condensate
$\langle qq(\mu)\rangle$ is that expected from Cooper-pairing at a Fermi
surface~\cite{Cotter:2012mb}, rather than the characteristic BEC form found in
$\chi$PT~\cite{Kogut:2000ek,Hands:2000ei}.

Beyond the fundamental requirements of determining the thermodynamic and
symmetry properties of the ground state, it is interesting to examine the
nature of excitations. As well as offering continuity with the traditional
concerns of lattice QCD at $T=\mu=0$, such questions bear on 
transport in the baryonic medium; answers to these questions in QCD would have the
potential to inform, say, descriptions of neutron star spin down (via
quantitative information on shear and bulk viscosities) and cooling (via a
knowledge of which if any excitations remain gapless and hence capable of carrying
energy away). 
There has been exploratory work in several directions. In~\cite{Hands:2007uc} the
hadron spectrum of QC$_2$D was calculated as a function of $\mu$; beyond $\mu_o$
in the meson sector the usual ordering $m_\pi<m_\rho$ is reversed, confirming
earlier studies~\cite{Muroya:2002ry}. Above onset the lightest states are found
in the $0^+$ and $1^+$ channels, with approximate degeneracy found between
mesons and diquarks, as might be expected in a superfluid phase in which baryon
number is no longer a good quantum number. The spectrum of heavy $QQ$
quarkonium states also shows a non-trivial $\mu$-dependence~\cite{Hands:2012yy},
possibly as a result of the formation of $Qq$ states in the quarkyonic regime.
In a recent study binding energies of multi-baryon ``nuclei''
formed from $0^+$ and $1^+$ bound states have been
estimated~\cite{Detmold:2014kba}.

On a different tack, quark and gluon propagators have been calculated
as functions of $T$ and $\mu$ in gauge-fixed
configurations~\cite{Hands:2006ve,Boz:2013rca}.
The electric (longitudinal) gluon propagator in Landau gauge becomes strongly
Debye-screened with increasing $T$  and $\mu$, whereas the magnetic
(transverse) gluon shows little sensitivity to $T$, and exhibits a mild
enhancement in the quarkyonic regime before becoming suppressed at large $\mu$.
Finally, the properties of topological excitations have been studied using a
cooling procedure to identify instantons~\cite{Hands:2011hd}. An enhancement of
topological susceptibility $\chi_T$ is seen on entering the quarkyonic regime,
which can be accommodated within the standard perturbative description of Debye
screening with the accompanying observation of a decrease in instanton scale
size $\rho(\mu)\propto\mu^{-2}$. $\chi_T$ does fall very steeply, however, once
$\langle L\rangle>0$.

In this work we attempt to probe the interaction between quarks,
and extract information on the spatial extent of hadrons, by calculating hadron
correlation functions in which the $q\bar q$ or $qq$ pair at the sink are
spatially separated by a vector $\vec r$~\cite{Velikson:1984qw}. 
For a bound state
$H$
whose temporal decay in Euclidean space is governed by a simple exponential
$e^{-E_Hx_0}$, the spatial profile, determined numerically as a function of
$\vec r$, is proportional to
the equal-time
Bethe-Salpeter wavefunction
\begin{equation}
\Psi(\vec r,\tau)=\int d^3\vec x\langle0\vert\bar\psi(\vec x,\tau)\psi(\vec x+\vec
r,\tau)\vert H\rangle.
\label{eq:wavefunction}
\end{equation}
The typical wavefunction profile for a bound state is gaussian, the width
giving basic information about the size of the hadron.
However, the correlators also yield interesting information even in the
absence of a bound state, as explored in a study of the Z$_2$ Gross-Neveu model
with $\mu\not=0$ in 2+1$d$~\cite{Hands:2003dh}. Above onset, the wavefunction 
is no longer positive definite, but rather has an oscillatory structure with
spatial frequency of order $k_F\sim\mu$. These oscillations have a similar
origin to the {\sl Friedel oscillations\/} observed in the density-density
correlations of electrons in metals (and thought to be responsible for the
spin-glass behaviour of certain alloys), characteristic of a sharp,
well-defined Fermi surface; the more primitive nature of the point-split hadron
correlator makes it easier to measure in a numerical simulation, however.
The observation of oscillatory wavefunctions in \cite{Hands:2003dh}, with
wavelength decreasing systematically with $\mu$, is one of several calculations
leading to the identification of the Z$_2$ GN model as a Fermi liquid.

A wavefunction study in QC$_2$D has the potential to shed light on several
outstanding issues in gauge theories at non-zero chemical potential, the most
fundamental being whether it is indeed possible to identify a well-defined Fermi
surface, since Fermi momentum $k_F$ is not a gauge-invariant quantity.  It may
also help clarify the nature of the quarkyonic state, which roughly speaking may
be thought of as a degenerate quark system in which only gauge-invariant
excitations are permitted. Since two-quark interactions are the most relevant at
a Fermi surface in the renormalisation group sense~\cite{Shankar:1993pf}, to
what extent lessons learned with $N_c=2$ can be generalised to QCD remains to be
seen. Nonetheless in principle the wavefunction should be a useful tool to chart
the passage from BEC to BCS realisations of superfluidity as $\mu$ increases,
which theoretically should take place for QC$_2$D near enough the chiral limit.
All these reasons motivate the current, exploratory study.

\section{Formulation}

In this section we explore the theoretical expectations for the wavefunction as
a function of interquark separation $r$. We begin, following
\cite{Hands:2003dh}, with the expression for the meson correlator $C_m(x_0;\vec
r)$ with a local point source at the origin, and $q$ and $\bar q$ separated by
$\vec r$ at the sink.  In anticipation of our later numerical results we choose
the {\it a priori\/} arbitrary sign of $\mu$ to yield the slowest decaying
result in the positive $x_0$ direction in diquark channels with non-zero baryon
charge.  Initially we assume free fields with quark mass $m$, and work at
strictly zero temperature; the chemical potential $\mu$ can then be understood
as a Fermi energy for a system of degenerate quarks with Fermi energy
$E_F(\mu)\equiv\mu=\sqrt{k_F^2+m^2}$. The onset value at which the ground state contains a
non-zero matter density is thus $\mu_o=m$: 
\begin{equation} C_m(x_0,\vec
r)=\sum_{\vec x}\mbox{tr}\int{d^4p\over(2\pi)^4}\int{d^4q\over(2\pi)^4}
\Gamma{{e^{ipx}}\over{ip{\!\!\! /}\,-\mu\gamma_0+m}} \Gamma{{e^{-iqx}e^{-i\vec
q.\vec r}}\over{iq{\!\!\! /}\,-\mu\gamma_0+m}}. \label{eq:start} 
\end{equation}
The Dirac matrix $\Gamma=\One,\gamma_5$ for channels $J^P=0^+,0^-$. 

Performing $\sum_{\vec x}$ and the trace over Dirac indices, we obtain
\begin{equation}
C_m^\pm(x_0;\vec r)={4\over{(2\pi)^5}}\int dq_0\int dp_0\int d^3\vec p
{{[\pm(p_0+i\mu)(q_0+i\mu)\pm\vec p^{\,2}+m^2]e^{i(p_0-q_0)x_0}e^{-i\vec p.\vec
r}}\over
{[(p_0+i\mu)^2+\vec p^{\,2}+m^2][(q_0+i\mu)^2+\vec p^{\,2}+m^2]}},
\label{eq:int2}
\end{equation}
where the $\pm$  sign denotes parity. First consider the 
case $0^-$. The denominator of (\ref{eq:int2}) has poles at $p_0,q_0=-i\mu\pm
i\sqrt{\vec p^{\,2}+m^2}$. Below onset, ie. for $\mu<m$, the integrals over $p_0$
and $q_0$ only yield a non-vanishing result if the $+$ root is chosen for
$p_0$ and the $-$ root for
$q_0$. We then use Cauchy's theorem to obtain
\begin{eqnarray}
C_m^-(x_0;\vec r) &=& {2\over{(2\pi)^3}}\int d^3\vec p e^{-2x_0\sqrt{\vec
p^2+m^2}}e^{-i\vec p.\vec r}\nonumber\\
&=&{2\over{(2\pi)^2}}\int_0^\infty dp p^2e^{-2x_0\sqrt{p^2+m^2}}\int_0^\pi
d\theta\sin\theta e^{ipr\cos\theta},
\end{eqnarray}
where in the last step $\vec r$ is taken to point towards the south pole.
The angular integral then yields
\begin{equation}
C_m^-(x_0;\vec r)={2\over{(2\pi)^{3\over2}}}r^{-{1\over2}}\int_0^\infty
dpp^{3\over2}e^{-2x_0\sqrt{p^2+m^2}}J_{1\over2}(pr),
\label{eq:angular}
\end{equation}
where  $J_{1\over2}(z)\equiv(2/\pi z)^{1\over2}\sin z$ is a Bessel function.
Now expand $(p^2+m^2)$ in
powers of $p/m$ and use Laplace's method of asymptotic
expansion:
\begin{eqnarray}
C_m^-(x_0;\vec
r)&\approx&{2\over{(2\pi)^{3\over2}}}r^{-{1\over2}}e^{-2x_0m}\int_0^\infty
dpp^{3\over2}e^{-{{p^2x_0}\over m}}J_{1\over2}(pr)\nonumber\\
&=&{{\surd\pi}\over{(2\pi)^2}}\left({m\over x_0}\right)^{3\over2}e^{-2mx_0}
\exp\left(-{{mr^2}\over{4x_0}}\right).
\label{eq:free_gaussian}
\end{eqnarray}
The corresponding expression
for $C_m^+$ has an extra factor $-p^2/(p^2+m^2)$ under the final $p$-integral,
so that $\vert C_m^-\vert>\vert C_m^+\vert$. The result
(\ref{eq:free_gaussian}) has
engineering
dimension 3 consistent with (\ref{eq:start}), is independent of $\mu$ as
expected below onset, and
decays in euclidean time with mass
$2m$, also as expected. The wavefunction profile is a Gaussian whose width
increases as $\sqrt x_0$, as appropriate for two free particles gradually drifting apart.

If instead we consider the corresponding diquark correlator
\begin{equation}
C_b(x_0,\vec r)=\sum_{\vec
x}\mbox{tr}\int{d^4p\over(2\pi)^4}\int{d^4q\over(2\pi)^4}
({\cal C}\Gamma){{e^{ipx}}\over{ip{\!\!\! /}\,-\mu\gamma_0+m}}
({\cal C}\Gamma)^{-1}\left({{e^{iqx}e^{i\vec q.\vec r}}\over{iq{\!\!\!
/}\,-\mu\gamma_0+m}}\right)^{tr},
\label{eq:diquark_start}
\end{equation}
where the charge conjugation matrix satisfies ${\cal C}=-{\cal C}^{-1}$ and
${\cal C}\gamma_\mu^{tr}{\cal C}^{-1}=-\gamma_\mu$, 
the result which emerges is
\begin{equation}
C_b^{\pm}(x_0,\vec r)=e^{2\mu x_0}C_m^\mp(x_0,\vec r).
\label{eq:d-m}
\end{equation}

\begin{figure}[t]
\begin{center}
    \includegraphics[width=0.5\textwidth,angle=-90]{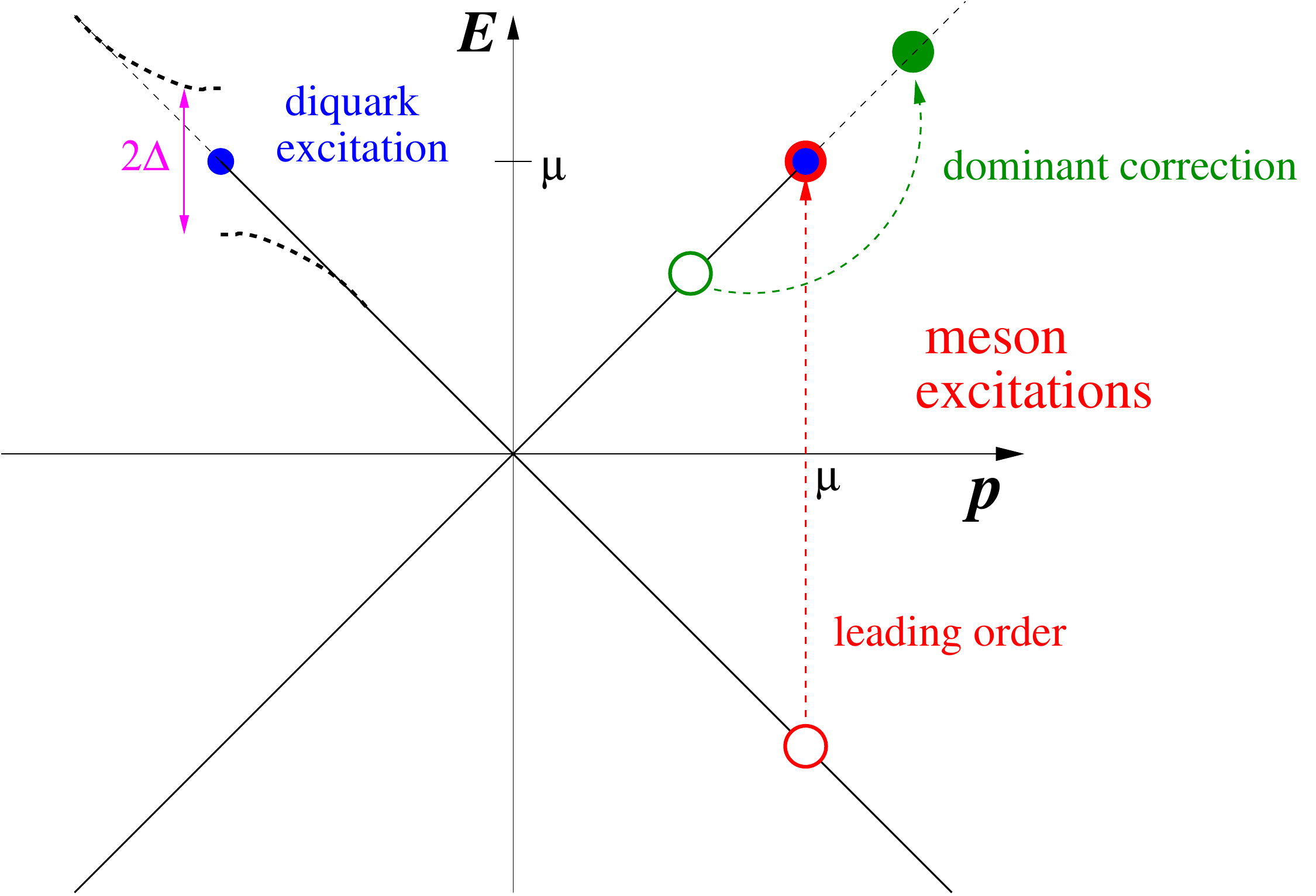}
\caption{Figure showing excitations of a system of massless
degenerate fermions in the zero total momentum sector: (a)  with meson quantum
numbers requiring energy $E=2\mu$ at leading order (red); (b)  with $E\ll\mu$ following one-gluon
exchange (green); and (c) with diquark quantum numbers
requiring zero energy (blue). Also shown in magenta is the
modifying effect of a gap $\Delta>0$ opening at the Fermi surface.}
\label{fig:excite}
\end{center}
\end{figure}

Next consider the more interesting case above onset $\mu>m$ (the value of
$\mu_o$ is governed by the mass per quark of the lightest baryon in the theory), where 
quark density $n_q=\langle\bar q\gamma_0 q\rangle>0$. For simplicity we set
$m=0$ so the only physical scale is $\mu$ (for a sufficiently large Fermi surface
this is always justified), and implying equality of $+$ and $-$ channels up
to a sign. 
Now the $p_0$ integral in (\ref{eq:int2})
can only receive a contribution from the pole at $-i(\mu-\vert\vec
p\vert)$ for $\vert\vec p\vert>\mu$, to yield
\begin{equation}
C_m(x_0;\vec r)={2\over{(2\pi)^{3\over2}}}r^{-{1\over2}}\int_\mu^\infty
dpp^{3\over2}e^{-2px_0}J_{1\over2}(pr).
\label{eq:lowerlimit}
\end{equation}
The only difference with (\ref{eq:angular}) is the lower limit of the
$p$-integral.
On changing integration variable $q=p-\mu$ and expressing 
the integrand as $e^{-2x_0(q+\mu)}A(q\mu)$ where $A$ can be expanded in powers
of $q$ using $J_\nu^\prime(z)=J_{\nu-1}(z)-(\nu/z)J_\nu(z)$, 
we find
\begin{equation}
C_m(x_0;\vec r)\approx{2\over{(2\pi)^{3\over2}}}{{e^{-2x_0\mu}\mu^{3\over2}}\over
r^{1\over2}}\int_0^\infty dqe^{-2x_0q}\left(1+{3\over2}{q\over\mu}\right)\left[
J_{1\over2}(r\mu)+qrJ^\prime_{1\over2}(r\mu)\right].
\end{equation}
In fact, the leading behaviour as $x_0\to\infty$ is given by the term of
$O(q^0)$:
\begin{eqnarray}
\lim_{x_0\to\infty}C_m(x_0;\vec
r)&=&{1\over{(2\pi)^{3\over2}}}{{\mu^{3\over2}e^{-2\mu x_0}}\over{x_0
r^{1\over2}}}J_{1\over2}(r\mu)\left(1+O(x_0^{-1})\right)\nonumber\\
&=&
{2\over{(2\pi)^2}}{\mu\over{x_0r}}e^{-2\mu
x_0}\sin(r\mu)\left(1+O(x_0^{-1})\right).
\label{eq:free_bessel}
\end{eqnarray}
The expression (\ref{eq:free_bessel}) also has the correct engineering dimension. This time the state
decays with energy $2\mu$, 
and arises from the promotion of a fermion from a negative energy state to the
Fermi surface 
by injection of  $\Delta E=2\mu$, $\Delta p=0$ 
(see Fig.~\ref{fig:excite}), leaving a negative energy hole which is
re-interpreted as an anti-quark.  
The wavefunction envelope now decays algebraically (ie.
$\propto r^{-1}$) rather
than exponentially as in (\ref{eq:free_gaussian}), and is modulated by 
oscillations of spatial frequency $\mu$. Remarkably, the wavefunction
can take negative values in this post-onset regime.
The relation (\ref{eq:d-m}) continues to be respected; 
diquarks can be interpreted as gapless excitations (ie. $\Delta E=0$) formed from bound quarks with
momenta $\pm p$ at the Fermi surface, as
indicated in Fig.~\ref{fig:excite}(c).

This is qualitatively similar to the
phenomenon known in many-body theory as {\sl Friedel Oscillations\/}, seen in the density-density
correlation function in degenerate systems. However, there
are important differences; strictly Friedel oscillations have spatial frequency
$2\mu$, and decay with a different power~\cite{Hands:2003dh}. Both phenomena can
be traced to the existence of a well-defined Fermi surface in the ground state
above onset, leading to a sharp lower limit on the integral in
(\ref{eq:lowerlimit}); the function $C(x_0,\vec r)$ is much easier to measure using lattice
techniques.  The oscillations were observed in simulations of a 2+1$d$
$N_f$-flavor four-fermion model in \cite{Hands:2003dh}; here the applicability of free
field theory could  be
justified by a calculation of the quantum corrections to the $q$ - $\bar q$
interaction in the large-$N_f$ limit. In the static (ie. $q_0=0$) limit this
interaction is totally screened, ie. the effective Debye mass $M(\mu)\to\infty$.

Since interactions between quarks clearly can't be ignored in QC$_2$D, a more
ambitious goal is to calculate $C(x_0;\vec r)$ in the presence of gauge interactions.
As a first step, we consider one-gluon exchange between massless fermions using the
Feynman-gauge propagator $\delta_{\mu\nu}F(q^2)$ (color factors are
ignored). A simple possibility for $F$ 
is the Debye-screened form $(q^2+M(\mu)^2)^{-1}$ where in a gauge theory the
expected relation is $M/\mu\sim O(g)$ where $g$ is the Yang-Mills coupling. The expression for $C_m$
is now a two-loop integral:
\begin{equation}
C_m=g^2\sum_{\vec x}\mbox{tr}\int_p\int_{\tilde p}\int_q F(q^2) \left\{
\Gamma{1\over{i(p{\!\!\! /}\,+q{\!\!\! /}\,)-\mu\gamma_0}}\gamma_\nu
{e^{ipx}\over{ip{\!\!\! /}\,-\mu\gamma_0}}\Gamma
{e^{-i\tilde p(x+\vec r)}\over{i\tilde p{\!\!\! /}\,-\mu\gamma_0}}\gamma_\nu
{1\over{i(\tilde p{\!\!\! /}\,+q{\!\!\! /}\,)-\mu\gamma_0}}\right\}.
\end{equation}
For simplicity's sake, consider the static limit $q_0=0$ and $\Gamma=\gamma_5$:
\begin{eqnarray}
&&C_m(x_0;\vec r)=-{16g^2\over{(2\pi)^9}}\int dp_0\int d\tilde p_0\int d^3\vec p\int d^3\vec q
e^{i(p_0-\tilde p_0)x_0}e^{-i\vec p.\vec r}F(\vec q^{\,2})\times\nonumber\\
&&{{[(p_0+i\mu)(\tilde p_0+i\mu)+\vec p^2][(p_0+i\mu)(\tilde p_0+i\mu)+(\vec
p+\vec q)^2]}\over
{[(p_0+i\mu)^2+(\vec p+\vec q)^2]
[(p_0+i\mu)^2+\vec p^2]
[(\tilde p_0+i\mu)^2+\vec p^2]
[(\tilde p_0+i\mu)^2+(\vec p+\vec q)^2]}}.
\label{eq:2loop}
\end{eqnarray}
    
The integrand of (\ref{eq:2loop}) has 4 poles $p_0=-i\mu\pm i\vert\vec
p+\vec q\vert$,
$p_0=-i\mu\pm i\vert\vec p\vert$, and 4 similar poles for $\tilde p_0$, yielding
in principle 16 cases to examine. However, the requirement
for the $\tilde p_0$ and $p_0$ poles to be in opposite half-planes
restricts us to only 6 non-vanishing possibilities.
For either $p_0=-i\mu+i\vert\vec p\vert$ or $p_0=-i\mu+i\vert\vec p+\vec q\vert$ we can have
$\tilde p_0=-i\mu-i\vert\vec p\vert$ or $\tilde p_0=-i\mu-i\vert\vec p+\vec q\vert$; there
will also be a contribution in a  restricted corner of phase space from 
$\tilde p_0=-i\mu+i\vert\vec p+\vec q\vert$. It turns out that the contributions
from the first two of the $\tilde p_0$ poles listed decay in Euclidean time at least as fast as
$e^{-2\mu x_0}$ and $e^{-\mu x_0}$ respectively, so we focus on the case
$p_0=-i\mu+i\vert\vec p\vert$, $\tilde p_0=-i\mu+i\vert\vec p+\vec q\vert$.
With this pole condition the integral becomes
\begin{equation}
C_m(x_0;\vec r)={4g^2\over{(2\pi)^7}}\int d^3\vec p \int d^3\vec q e^{-(\vert\vec p\vert-\vert\vec
p+\vec q\vert)x_0}e^{-i\vec p.\vec r}{F(q^2)\over{(\vert\vec p\vert+\vert\vec
p+\vec q\vert)^2}},
\end{equation}
with the range of integration restricted to $\vert\vec
p\vert>\mu$, $\vert\vec p+\vec q\vert<\mu$, implying $\vert\vec p\vert-\vert\vec
p+\vec q\vert>0$. 
There are regions of phase space,
with $\vert\vec q\vert/\vert\vec p\vert\approx O(1)$, 
where the integrand falls away much 
more slowly than $e^{-\mu x_0}$. These regions are due to the excitation of a
quark to a state lying above the Fermi surface leaving a hole in a positive
energy state, with energy $\Delta E\ll\mu$ in general, which is now permitted by the
kinematics with $\Delta p=q\not=0$ (see Fig.~\ref{fig:excite}(b)).

Our tactic will be to perform $\int d^3\vec q$ first. The pole conditions
dictate that 
the angle between $\vec q$ and $-\vec p$
is constrained to $\theta_q<\Theta(p)=\sin^{-1}{\mu\over p}$, and that 
the magnitude $q$ is restricted to lie between
$p\cos\theta_q(1\pm\sqrt{1-\cos^2\Theta/\cos^2\theta_q})$
(for $\theta_q=0$ this reduces to $q\in(p-\mu,p+\mu)$).
We arrive at
\begin{eqnarray}
&&{4g^2\over{(2\pi)^5}}\int_\mu^\infty p^2 dp\int_0^\pi d\theta_p\sin\theta_p
e^{ipr\cos\theta_p}\int_0^{\Theta(p)}d\theta_q\sin\theta_q \nonumber\\
&\times&
\int_{q_{min}(p,\theta_q)}^{q_{max}(p,\theta_q)}dq
{{q^2e^{-(p-\sqrt{p^2+q^2-2pq\cos\theta_q})x_0}}\over
{(q^2+M^2)(2p^2+q^2-2pq\cos\theta_q+2p\sqrt{p^2+q^2-2pq\cos\theta_q})}}\nonumber\\
&\equiv&{4g^2\over{(2\pi)^{9\over2}}}r^{-{1\over2}}\int_\mu^\infty
dpp^{3\over2}J_{1\over2}(pr)I(p,x_0),\label{eq:cont3}
\end{eqnarray}
where the Debye-screened form of $F(q^2)$ has been inserted.

To examine things more closely,
consider the limit $\theta_q=0$ so that $\vec p+\vec q$ is either
parallel or anti-parallel to $\vec p$. The integral over $q$ in (\ref{eq:cont3})
then becomes
\begin{equation}
\int_{p-\mu}^p dq {{q^2e^{-qx_0}}\over{(q^2+M^2)(2p-q)^2}}+\int_p^{p+\mu} dq
{{e^{-qx_0}}\over{q^2+M^2}}.
\label{eq:lastbit}
\end{equation}
After some rearrangement the integrals may be performed to yield
\begin{equation}
{1\over x_0}\left(-{{e^{-(p+\mu)x_0}}\over{(p+\mu)^2+M^2}}+e^{-px_0}\times0+
{{e^{-(p-\mu)x_0}(p-\mu)^2}\over{(p+\mu)^2[(p-\mu)^2+M^2]}}\right)(1+O(x_0^{-1}).
\end{equation}
Reassuringly the factor multiplying
$e^{-px_0}$ cancels at this order in $x_0^{-1}$; this dependence could only have arisen
via incomplete cancellation between the inner limits on the integrals in
(\ref{eq:lastbit}), but the behaviour of the integrand at the origin of
$p$-space should be smooth. Leaving aside for now  the slightly ill-defined integral over
$\theta_q$, we arrive at an approximate expression of the form 
\begin{eqnarray}
C_m(x_0,\vec r)&\sim&{4g^2\over{x_0(2\pi)^5}}\int_\mu^\infty p^2dp\int_0^\pi d\theta_p \sin\theta_p
e^{ipr\cos\theta_p}
{{e^{-(p-\mu)x_0}(p-\mu)^2}\over{(p+\mu)^2[(p-\mu)^2+M^2]}}\nonumber\\
&=&{4g^2\over{(2\pi)^{9\over2}x_0r^{1\over2}}}\int_\mu^\infty dp p^{3\over2}
J_{1\over2}(pr){(p-\mu)^2\over(p+\mu)^2}{{e^{-(p-\mu)x_0}}\over{(p-\mu)^2+M^2}},
\end{eqnarray}
where terms decaying faster than $e^{-2\mu x_0}$ and/or $x_0^{-2}$ are
neglected, and in the second line the integral over $\theta_p$ has been
completed.

Finally, as in the one-loop case we substitute $q=p-\mu$ and
Taylor-expand the integrand in powers of $q$. The identity
$\int_0^\infty q^n e^{-qx_0}dq=n!/x_0^{n+1}$ is exploited to yield
\begin{equation}
C_m(x_0,\vec r)\sim {2g^2\over{(2\pi)^{9\over2}}}{{J_{1\over2}(\mu r)}\over{(\mu r)^{1\over2}}}
{1\over{x_0^4M^2}}(1+O(x_0^{-1}))
\sim {{x_0^{-2}}\over{(x_0\mu)^2}}{{J_{1\over2}(r\mu)}\over{(r\mu)^{1\over2}}}.
\label{eq:friedel_int}
\end{equation}
The engineering dimension is wrong due to the approximation that has been carried out preceding
eqn.~(\ref{eq:lastbit}). There is no exponential
decay in $x_0$, resulting from the presence in the integrand of poles costing zero
excitation energy, whose precise contribution depends on complicated geometrical
factors which we have simplified.  However the oscillatory
$r$-dependence, the main focus of this work, remains unchanged from the free field result
(\ref{eq:free_bessel}).  This form is dictated by the
final momentum integral always having  a sharp lower limit
$p=\mu$, arising from the physical situation of gapless excitations at a Fermi
surface. Note also that
the first correction  (\ref{eq:friedel_int}) can
potentially exceed the free-field result (\ref{eq:free_bessel}) for large
$\mu x_0$, highlighting
for the need for caution when interpreting perturbative estimates.

In the next section the dependence of the wavefunction on a diquark source
strength $j$ will be investigated numerically using free lattice fermions. The
term proportional to $j$ in the action explicitly breaks baryon-number symmetry,
and encourages diquark pairing which has the effect of opening up a gap at the
Fermi surface; we will see that the impact on the oscillations is profound.

\section{Lattice Formulation}

The lattice results in this study employ two flavors of Wilson fermion, with
action for the quarks given by~\cite{Hands:2006ve}
\begin{equation}
S=\sum_{x,y}\sum_{i=1,2}\bar\psi_{ix}M_{xy}(\mu)\psi_{iy}-\kappa j\delta_{xy}[
-\bar\psi_{1x}({\cal C}\gamma_5)\tau_2\bar\psi_{2y}^T+\psi_{2x}^T({\cal
C}\gamma_5)\tau_2\psi_{1y}],
\label{eq:latt_act}
\end{equation}
with (in units where lattice spacing $a=1$)
\begin{equation}
M_{xy}(\mu)=\delta_{xy}-\kappa\sum_\nu\left[
(1-\gamma_\nu)e^{\mu\delta_{\nu
0}}U_\nu(x)\delta_{y,x+\hat\nu}+(1+\gamma_\nu)e^{-\mu\delta_{\nu
0}}U_\nu^\dagger(y)\delta_{y,x-\hat\nu}\right].
\label{eq:wilsonM}
\end{equation}
The subscripts on the quark fields denote flavor, and the Pauli matrix $\tau_2$
acts on color indices. The usual action is supplemented by the addition of gauge
invariant scalar isoscalar diquark source terms with strengths $j$, here chosen
real (gauge invariance follows from the SU(2) identity $\tau_2U\tau_2=U^*$).
These terms serve a dual role; they enable the extraction of diquark condensates
and anomalous quark propagators of the generic form $\langle
\psi(x)\psi^T(y)\rangle$, $\langle\bar\psi^T(x)\bar\psi(y)\rangle$, and their
inclusion also mitigates the long-wavelength fluctuations due to Goldstone modes
in the superfluid phase which forms for $\mu\geq\mu_o={1\over2}m_\pi$ in which baryon
number symmetry is spontaneously broken. As a direct consequence the number of
iterations required to invert the matrix $M$ is reduced by the opening of a gap
in the Dirac spectrum of $O(j)$. In order to make contact with a theory where
baryon number is a conserved charge we need ultimately to examine the limit
$j\to0$.

\begin{table}[t]
\begin{center}
    \includegraphics{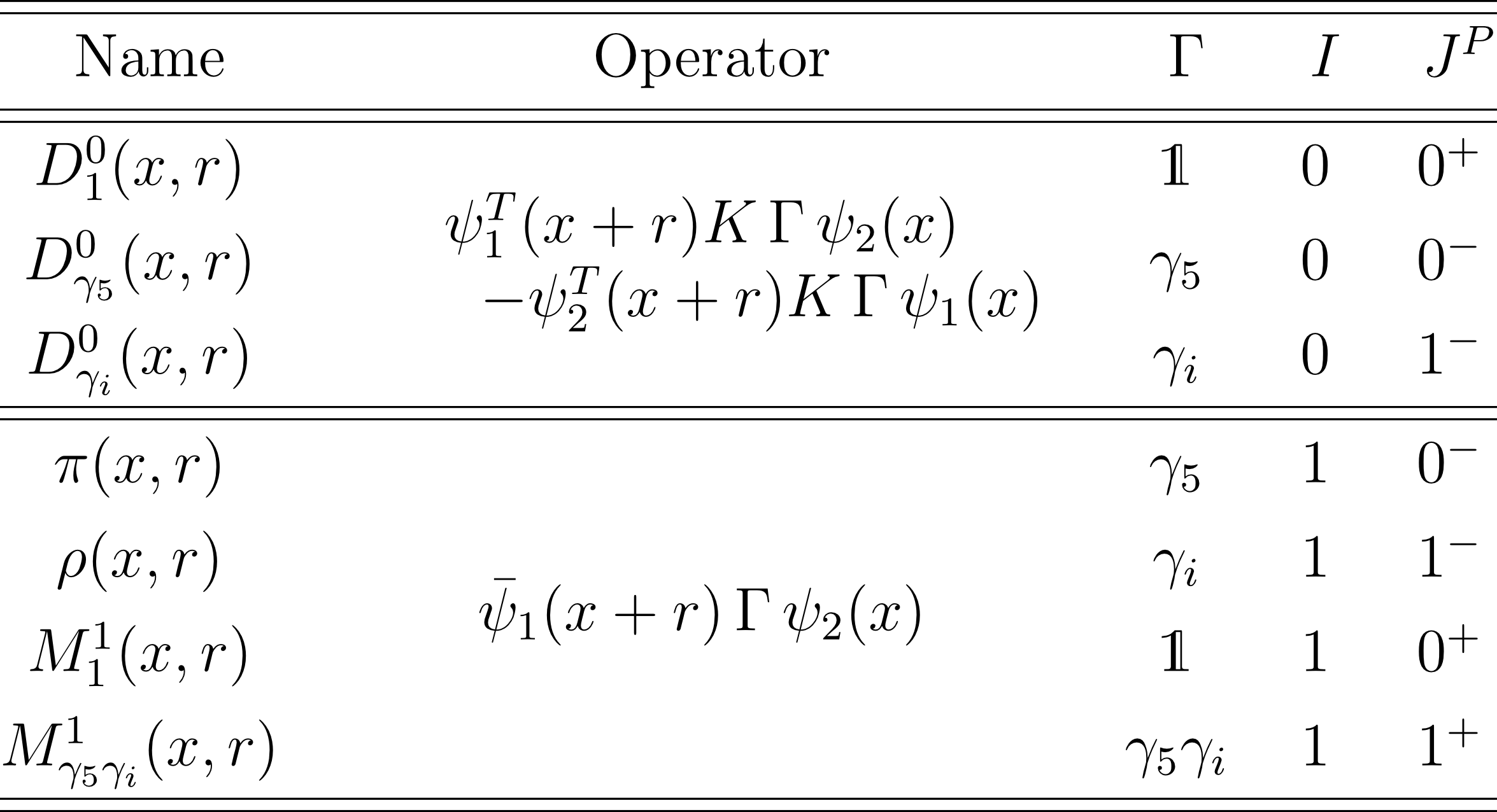}
\caption{A summary of the diquark and meson operators used in the analysis, with $ K={\cal C}\gamma_5\tau_2 $.} 
\label{fig:table}
\end{center}
\end{table}

The action (\ref{eq:latt_act}) may be rewritten
\begin{equation}
S=(\bar\psi,\bar\phi)
\left(
\begin{matrix}M(\mu) & \kappa j\gamma_5 \\ 
-\kappa j\gamma_5 & M(-\mu)\end{matrix}
\right)
\left(\begin{matrix}\psi \\ \phi\end{matrix}\right)
\equiv\bar\Psi{\cal M}(\mu,j)\Psi,
\label{eq:calM}
\end{equation}
using the relabellings $\psi=\psi_1$, $\bar\psi=\bar\psi_1$, $\phi=-{\cal
C}\tau_2\bar\psi_2^T$, $\bar\phi=-\psi_2^T{\cal C}\tau_2$.
The enhanced matrix ${\cal M}$ obeys an important identity
\begin{equation}
\Gamma_5{\cal M}(\mu,j)\Gamma_5={\cal M}^\dagger(-\mu,-j),
\label{eq:Gamma5}
\end{equation}
where $\Gamma_5\equiv\mbox{diag}(\gamma_5,\gamma_5)$. 
We can now outline hadron
interpolation operators, following the treatment in \cite{Hands:2007uc}. 
In this study we will consider isovector meson excitations of the form
\begin{equation}
M^1_\Gamma=\bar\psi_1\Gamma\psi_2=\bar\psi\Gamma({\cal C}\tau_2)\bar\phi^T
\end{equation}
and baryonic diquarks in both isoscalar
\begin{equation}
D^0_\Gamma=\psi_1^T({\cal
C}\gamma_5\tau_2)\Gamma\psi_2=\psi^T\gamma_5\Gamma^T\bar\phi^T=-\bar\phi\Gamma\gamma_5\psi
\end{equation}
and isovector
\begin{equation}
D^1_\Gamma=\psi^T({\cal C}\gamma_5\tau_2)\Gamma\psi
\end{equation}
channels. The Dirac matrices $\Gamma$ 
and corresponding 
spacetime quantum numbers $J^P$ are listed in Table~\ref{fig:table}. 

The general quark propagator is
\begin{equation}
\langle\Psi\bar\Psi\rangle={\cal M}^{-1}(\mu,j)\equiv
\left(\begin{matrix}
S_{\psi\bar\psi}&S_{\psi\bar\phi}\\
S_{\phi\bar\psi}&S_{\phi\bar\phi}
\end{matrix}\right).
\end{equation}
The relations (\ref{eq:calM},\ref{eq:Gamma5}) imply two distinct components:
\begin{eqnarray}
\mbox{Normal:}&&S_{\psi\bar\psi}(\mu,j)=M^{-1}(\mu,j)=S_{\phi\bar\phi}(-\mu,-j);\\
\mbox{Anomalous:}&&S_{\psi\bar\phi}(\mu,j)=S_{\phi\bar\psi}(-\mu,-j).
\end{eqnarray}
In the limit $j\to0$ and on a finite lattice the anomalous components vanish.
We can now write expressions for the hadron correlators required for the
wavefunctions corresponding to spatial separation $\vec r$ between the
(anti)quarks at the sink. First define eg. 
$M^1_\Gamma(x,\vec r)=\bar\psi_1(x)
\Gamma\psi_2(x+\vec r)$, and $M^{1\dagger}_\Gamma=-\bar\psi_2(x+\vec
r)\bar\Gamma\psi_1(x)$ (with $\bar\Gamma\equiv\gamma_0\Gamma^\dagger\gamma_0$).
Then 
\begin{eqnarray}
C_{m\Gamma}^1(x,y;\vec r)=\langle M^1_\Gamma(y,\vec0)M_\Gamma^{1\dagger}(x,\vec
r)\rangle
=&-&\mbox{tr}\bigl\{S_{\psi\bar\psi}(x,y)\Gamma
\gamma_5S_{\phi\bar\phi}^\dagger(x+\vec r,y)\gamma_5\bar\Gamma\bigr\}\nonumber\\
&+&\mbox{tr}\bigl\{S_{\phi\bar\psi}(x+\vec r,y)\Gamma\gamma_5
S_{\psi\bar\phi}^\dagger(y,
x)\gamma_5\bar\Gamma\bigr\}.
\label{eq:isovector_meson}
\end{eqnarray}
There are two terms, one involving a connected contraction of normal
propagators, and one a connected contraction of anomalous propagators. 

Similarly in the isoscalar diquark channel (with $D=\psi_1^{T}({\cal
C}\gamma_5\tau_2)\Gamma\psi_2$ and $D^\dagger=\bar\psi_2\bar\Gamma({\cal
C}\gamma_5\tau_2)\bar\psi_1^T$):
\begin{eqnarray}
C_{b\Gamma}^0(x,y;\vec r)=\langle D^0_\Gamma(y,\vec0))D_\Gamma^{0\dagger}(x,\vec
r)\rangle
=&+&\mbox{tr}\bigl\{S_{\psi\bar\phi}(y,y)\Gamma\gamma_5\bigr\}
\mbox{tr}\bigl\{S_{\phi\bar\psi}(x,x+\vec r)\bar\Gamma\gamma_5\bigr\}\nonumber\\
&-&\mbox{tr}\bigl\{S_{\phi\bar\phi}(x,y)\Gamma S_{\phi\bar\phi}^\dagger(x+\vec r,
y)\bar\Gamma\bigr\}.
\label{eq:isoscalar_diquark}
\end{eqnarray}
Here the anomalous
propagators only contribute to disconnected pieces, which we henceforth
ignore. A judicious use of (\ref{eq:Gamma5}) means the required normal
contributions can be evaluated with a single inversion of ${\cal M}$ using a
local source at $y$, just as in (\ref{eq:isovector_meson}). It is shown in \cite{Hands:2007uc} that the connected
contribution to $C^0_{b\Gamma}$ is non-vanishing for
$\Gamma\in\{\One,\gamma_5,\gamma_i\}$. Comparison of (\ref{eq:isovector_meson})
and (\ref{eq:isoscalar_diquark}) also reveals the origin, in the limit
$\mu=j=0$,
of the degeneracy between mesons with spin-parity $J^P$ and diquarks with
$J^{-P}$, noted in \cite{Hands:2007uc}. Once the system enters a
superfluid phase for $\mu\geq\mu_o$ baryon number is no longer a good quantum
number, and we therefore expect degeneracy between meson and diquark
channels sharing the same $J^P$. This prediction was approximately observed 
for the lightest states in \cite{Hands:2007uc}, and must arise via
non-trivial contributions from anomalous propagators. Finally, we note that a
similar  analysis for the isovector
diquark correlator $C^1_{b\Gamma}$ yields 
the same connected contribution as in (\ref{eq:isoscalar_diquark}), 
but this time with $\Gamma$ restricted to $i\gamma_5\gamma_i$.
 
\begin{figure}[t]
\begin{center}
    \includegraphics{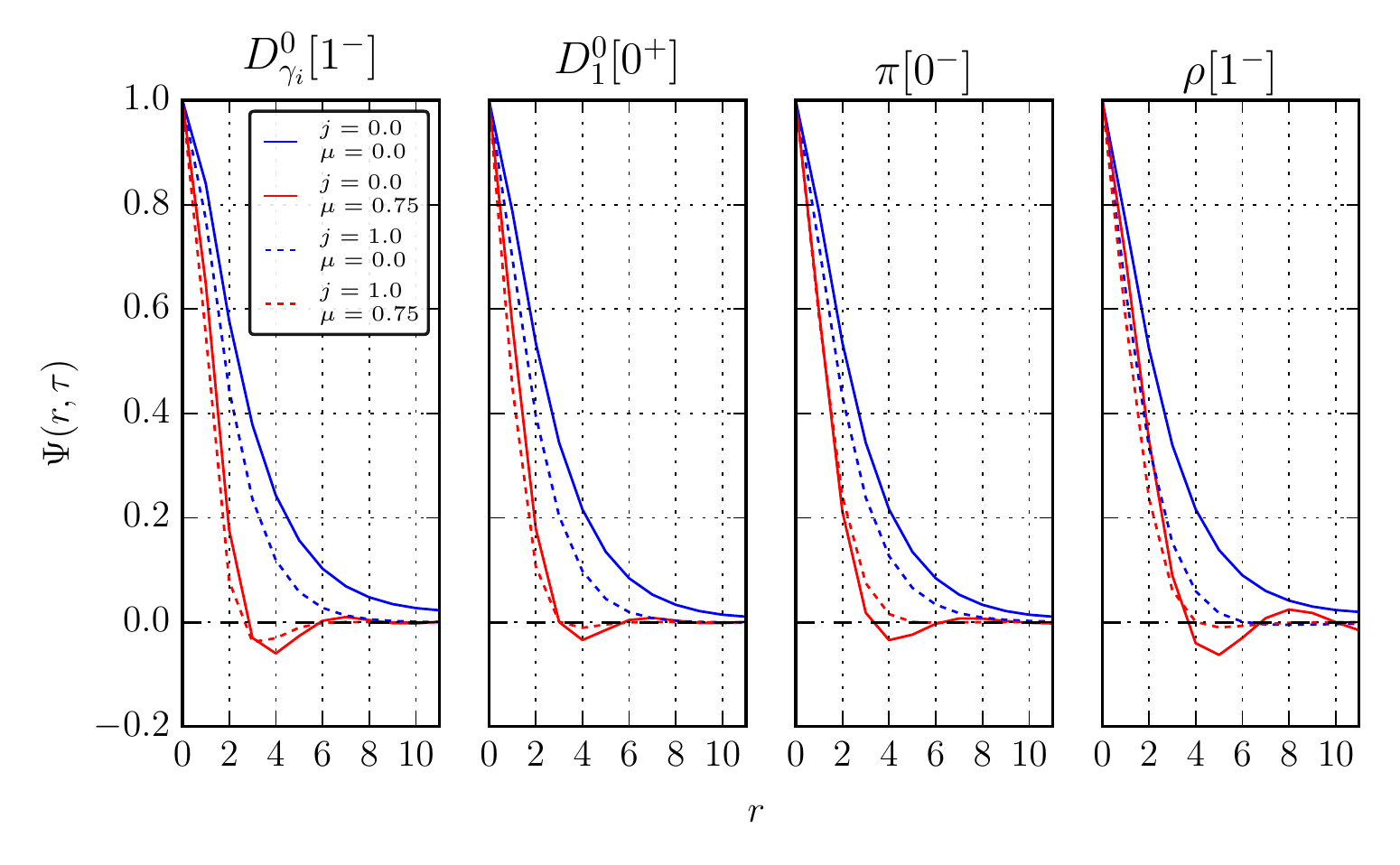}
\caption{Meson and diquark wavefunctions $\Psi(\vec r,\tau=8)$ 
for free massless quarks and various $\mu$ and $j$.} 
\label{fig:free}
\end{center}
\end{figure}
Next we examine wavefunctions evaluated for non-interacting quarks.
Fig.~\ref{fig:free} plots the wavefunction $\Psi_\Gamma^1(\vec r,\tau)$ defined by the
ratio
\begin{equation}
\Psi(\vec r,\tau)={{\sum_{\vec x}C(\vec x,\tau;0;\vec r)}\over{\sum_{\vec x}C(\vec
x,\tau;0;\vec0)}},
\label{eq:ratio}
\end{equation}
where as usual the sum over the sink location $\vec x$ projects onto the zero
momentum sector. The calculation is performed by inverting ${\cal M}$ on a $24^4$
lattice, with $\kappa={1\over8}$ chosen to yield massless quarks. For
consistency's sake all curves shown correspond to temporal separation $\tau=8$
between source and sink, and only points evaluated along a lattice
axis are plotted to minimise the impact of discretisation artifacts.

Two trends are discernible. First, as
chemical potential increases from 0 to $\mu a=0.75$ the width of the main central feature of the
free wavefunction decreases, and it develops an oscillatory
structure whose spatial
frequency is
in qualitative agreement (but slightly smaller) than the prediction of (\ref{eq:free_bessel}). 
The amplitude of the oscillations dies
away more quickly than for those observed in the $2+1$d Gross-Neveu model
\cite{Hands:2003dh}, since in $3+1$d the profile is given by
$J_{1\over2}(r\mu)$ rather than $J_0(r\mu)$. 
Second, as diquark source $j$ is increased at fixed $\mu$, both the width of the
central feature (especially for $\mu=0$) and, for $\mu>0$  the amplitude of the oscillations are diminished. 
This can be understood as a consequence of the opening of a gap at the Fermi surface over
a momentum range $\Delta\sim O(j)$, as shown in Fig.~\ref{fig:excite}, with the
consequence that there is mixing between particle and hole states so that both higher
momentum states contribute to the correlators, and the sharp momentum cutoff
in the integral leading to (\ref{eq:free_bessel}) is smeared out. 

\section{Numerical Results}
\label{sec:results}
Hadron wavefunctions formed from interacting quarks were calculated using
QC$_2$D ensembles generated using the quark action
(\ref{eq:latt_act},\ref{eq:wilsonM}) together with an unimproved Wilson gauge
action for the gluons. The simulation parameters were $\beta=1.9$,
$\kappa=0.168$, corresponding to lattice spacing $a=0.178(5)$fm with scale
set by assuming the string tension is (440MeV)$^2$, and mass ratio
$m_\pi/m_\rho=0.807(5)$~\cite{Cotter:2012mb}. Most results are obtained on a
$12^3\times 24$ lattice, corresponding to a physical temperature $T=44(2)$MeV,
although for $\mu a=0.25$ we also have results from $16^3\times24$ for
comparison. 
This temperature is sufficiently low to support the existence of an extended
range of $\mu$ in which the theory is simultaneously confining (as indicated by
a near-vanishing Polyakov loop) and superfluid (as indicated by a non-vanishing
condensate $\langle\psi^T_2({\cal C}\gamma_5)\tau_2\psi_1\rangle\not=0$ as
$j\to0$). With these parameters values of $\mu a$ in the range [0.0,1.1] were
explored; the onset value $\mu_o={1\over2}m_\pi=0.323(3)a^{-1}$. The so-called
``quarkyonic'' regime where baryon density, pressure and superfluid condensate all
scale with $\mu$ according to the expectations of a system of degenerate
quarks, lies approximately in the range $\mu
a\in(0.4,0.8)$~\cite{Cotter:2012mb,Cotter:2013oaa}.

For $\vert\vec r\vert>0$ the point-split correlators
(\ref{eq:isovector_meson},\ref{eq:isoscalar_diquark}) are not gauge invariant
without an insertion of path-ordered link variables along some selection of paths joining the two halves of the sink. 
To mitigate the effects of the signal fluctuations introduced by this non-unique procedure,
we instead choose to gauge-fix the configuration and use unit links to complete
the loop. We fix a discretised Coulomb gauge defined by
\begin{equation}
\Delta^G(x) \equiv \sum_{i=1}^3\left[A_i^G(x)-A_i^G(x-\hat\imath)\right]=0\,,
\end{equation}
where the gauge transformation $G(x)$ extremises the functional
\begin{equation}\label{eq:effel}
F[U^G]=-\mbox{Re}\,\mbox{Tr}\sum_x\sum_{i=1}^3U_i^{G}(x),
\end{equation}
with $U_\mu^{G }(x)=G(x) U_\mu(x) G^{-1}(x+\mu) $.
To achieve this, the simplest algorithm~\cite{Giusti:2001xf} one can adopt is a
local  procedure which visits one lattice site at a time and attempts to
minimize its contribution to the functional (\ref{eq:effel}), which can be
written as: \begin{equation} F_{\rm loc}(\bar x)\propto -{\rm Re}\, {\rm Tr}
\sum_\mu \left[ U_\mu (\bar x)+U_\mu(\bar x - \hat{\mu})\right]\,.
\end{equation} Two observables are usually monitored during this procedure. One
is the functional (\ref{eq:effel}) itself, which decreases monotonically and
eventually reaches a plateau. The other one is a measure of the first derivative
of $F[U]$ during the gauge-fixing process defined as  
\begin{equation}
\theta^G \equiv {\frac 1 V} \ \sum_{x}\mbox{Tr} \ [ \Delta^G (x) (\Delta^G)^{\dagger}
(x)]\,,
\label{eq:thetalat} 
\end{equation} 
where $V$ is the lattice volume. This
quantity eventually approaches zero when $F[U]$ reaches its minimum and can be
used as a stopping parameter for the procedure.  Here we chose $\theta \le
10^{-30}$.

As shown in~\cite{Giusti:2001xf}, the gauge-fixing procedure is affected by long-range correlations when  the size of the lattice increases, which eventually leads to a  critical slowing down. There are a number of strategies available to improve this and here we adopt the so-called  overrelaxation procedure~\cite{Mandula:1990vs}. This method replaces the gauge fixing transformation $G(x)$ with its power $G^{\omega}(x)$ at each iteration, where $\omega \in (1,2) $ is a parameter which is tuned empirically at an optimal value,  depending on the volume of the
lattice and the coupling $\beta$. 
The exponentiation is obtained by a truncated binomial expansion
\begin{equation}  
G^{\omega} = \sum_{n=0}^N \frac{\gamma_n(\omega)}{n!}{(G-I)}^n \, ,
\quad
\gamma_n(\omega) = \frac{\Gamma(\omega + 1)}{\Gamma(\omega + 1 - n)}\,,
\label{eq:o3}
\end{equation}
with $2\le N \le 4$. Here  we chose $\omega = 1.75$ and $N=4$. Before $G^{\omega}(x)$ is applied
to the link, it has to be reunitarised in order to belong to the gauge group. 

\begin{figure}[t]
\begin{center}
    \includegraphics{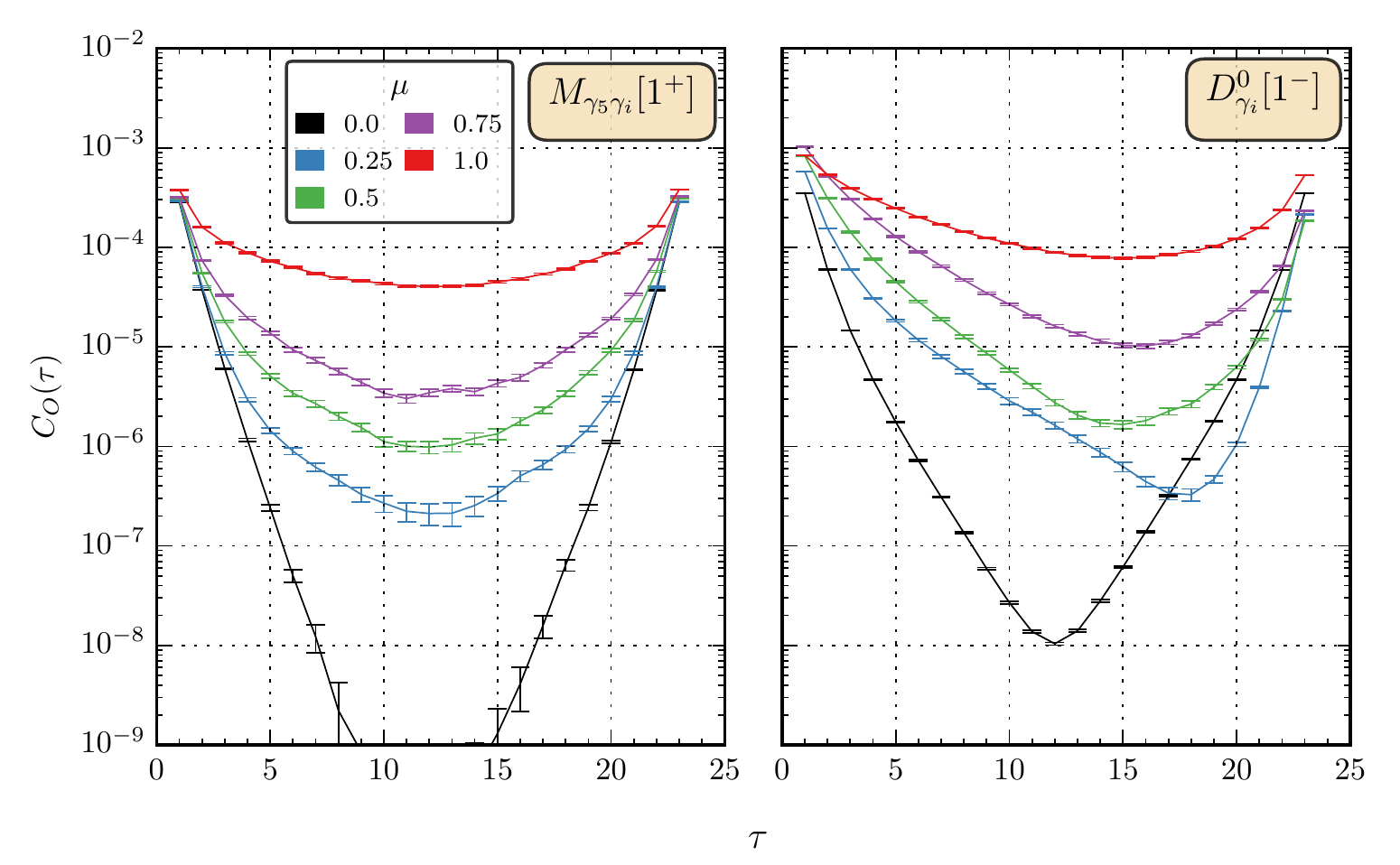}
\caption{Meson (left) and diquark (right) Euclidean correlators evaluated for various $\mu$ at  $ja=0.02$.}
\label{fig:corr}
\end{center}
\end{figure}
Fig.~\ref{fig:corr} shows typical meson and diquark timeslice correlators $\sum_{\vec
x}C_{m,b}(0,x;\vec r=\vec0)$ evaluated for $\mu a\in[0,1.0)$. By construction
mesons yield a signal which is symmetric about the temporal midpoint of the
lattice, whereas diquarks carrying baryon charge are time-asymmetric once $\mu>0$. We
interpret the forwards-moving lighter states as diquarks and the
backwards-moving heavier states as anti-diquarks (in contradistinction to the
nomenclature adopted in \cite{Hands:2007uc}). Inspection of the plots reveals
that while the $1^+$ meson is more massive than the $1^-$ diquark at $\mu=0$
(recall they should only be degenerate in the limit $j\to0$),
both become lighter as $\mu$ increases beyond onset. 

This is in accordance with the spectrum results found in \cite{Hands:2007uc} using
all-to-all propagators evaluated using source dilution. In brief, in mesonic channels, the 
$0^-$ pion and $1^-$ rho are the lightest states at $\mu=0$, but beyond onset
there is a level crossing, the rho becoming lighter while the pion mass
increases almost linearly with $\mu$. The two lightest states for $\mu>\mu_o$
are, however, the isoscalar $0^+$ and the slightly heavier isovector $1^+$. 
In the diquark channel, the isoscalar $0^+$ and isovector $1^+$ are
approximately degenerate with the pion and rho states at $\mu=0$ but, again,  beyond
onset become much lighter and approximately degenerate with the mesonic states
sharing the same quantum numbers, as expected in a superfluid ground state. 
In the same post-onset regime isoscalar $0^-$ and $1^-$ states are observed at
roughly the same mass scales as their  $\pi/\rho$ counterparts. 

The main focus of this paper is the post-onset regime $\mu>\mu_o$, where we
expect degeneracy between meson and baryon states due to the superfluid nature
of the ground state, so that physical states are labeled solely by their $J^P$
quantum numbers. However, meson and baryon operators need not have the same
overlap with the actual bound states;
Table 2 of Ref.~\cite{Hands:2007uc} shows the quality of the signal for each
operator.  We will mainly concentrate on the channels yielding a
``good'' signal for $\mu>\mu_o$: the $0^+$, $0^-$ and $1^-$ diquarks and the
$1^+$ meson. 

By $\tau=8$ the temporal decay
in both cases is well-approximated over a wide range of $\tau$ by a single exponential, indicative of a
bound state.  In this regime the ratio (\ref{eq:ratio}) can be written eg.
\begin{equation}
\Psi(\vec r,\tau)={{
\sum_i\sum_{\vec x}\langle
0\vert\bar\psi(0)\Gamma\psi(0)\vert i\rangle\langle
i\vert\bar\psi(x)\Gamma\psi(x+\vec r)\vert 0\rangle e^{-E_i x_0}}\over
{\sum_{i}\sum_{\vec x}\langle
0\vert\bar\psi(0)\Gamma\psi(0)\vert i\rangle\langle
i\vert\bar\psi(x)\Gamma\psi(x)\vert 0\rangle e^{-E_i x_0}}}\approx{
{\sum_{\vec x}\langle
H\vert\bar\psi(\vec x)\Gamma\psi(\vec x+\vec r)\vert 0\rangle}\over
{\sum_{\vec x}\langle
H\vert\bar\psi(\vec x)\Gamma\psi(\vec x)\vert 0\rangle}},
\label{eq:wvfn}
\end{equation}
where $\vert H\rangle$ is understood as the
lightest hadron state in the particular channel, 
and can be seen to be proportional to the wavefunction defined in
eqn.~(\ref{eq:wavefunction}) 
but normalised to unity at the
origin $\vec r =\vec 0$.
Unless otherwise stated, we will show results for wavefunctions
$\Psi$ evaluated at this temporal separation.

\begin{figure}[!t]
\begin{center}
    \includegraphics{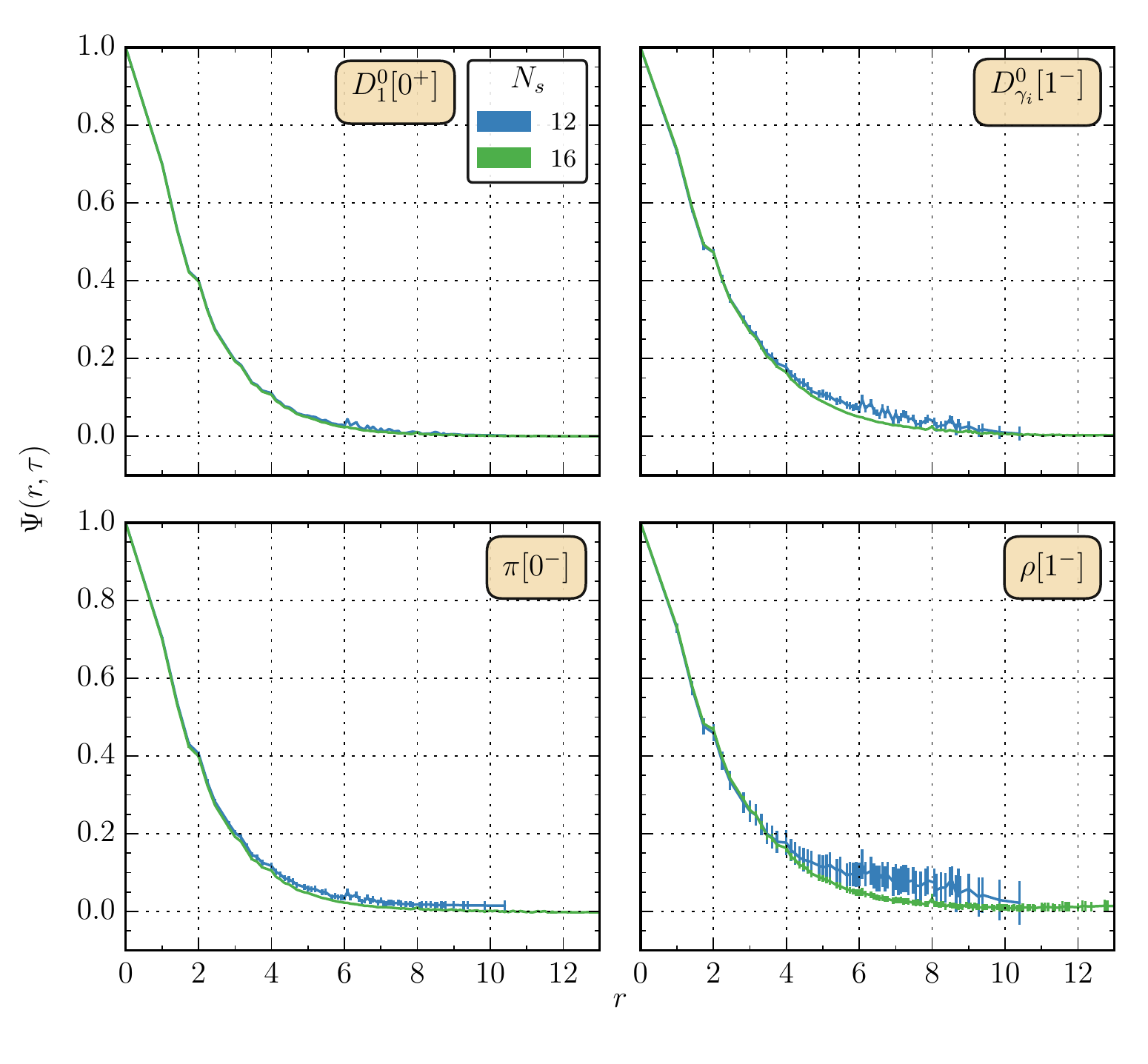}
\caption{Hadron timeslice correlators evaluated for $ja=0.02$ and $\mu a=0.25$ for various
channels.}
\label{fig:voldep}
\end{center}
\end{figure}
Fig.~\ref{fig:voldep} shows wavefunctions $\Psi(r,\tau)$ evaluated on timeslice 8 for the
same four channels of Fig~\ref{fig:free}. The plots compare data from
$12^3\times24$ and $16^3\times24$ lattices; data from all available separations
$\vec r$ up to half the lattice extent is shown.  There is some evidence for
discretisation artifacts, eg. around $r\simeq2a$, but overall the trend with $r$
is smooth within errors;
moreover finite volume effects
are absent for spin-0 states, and larger (but still consistent with zero) for
spin-1, where data from the smaller volume are considerably noisier. 
The absence of spatial volume effects is consistent with excitations being
confined states, with little or no contamination from image charges which might
be anticipated from the algebraic fall off (\ref{eq:free_bessel}) of the
free-field wavefunctions.

%
%
\begin{figure}[p]
\begin{center}
    \includegraphics{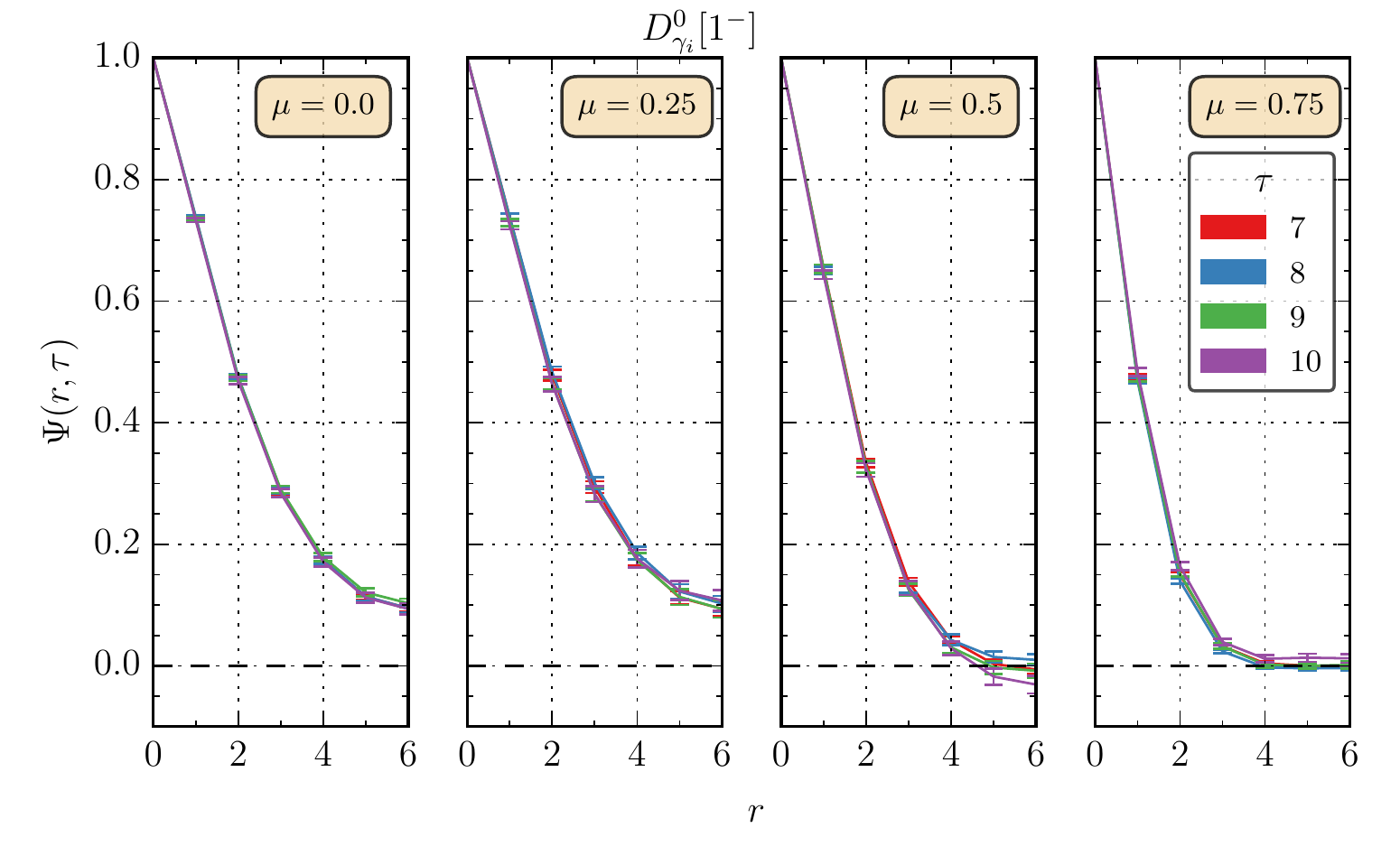}
    \includegraphics{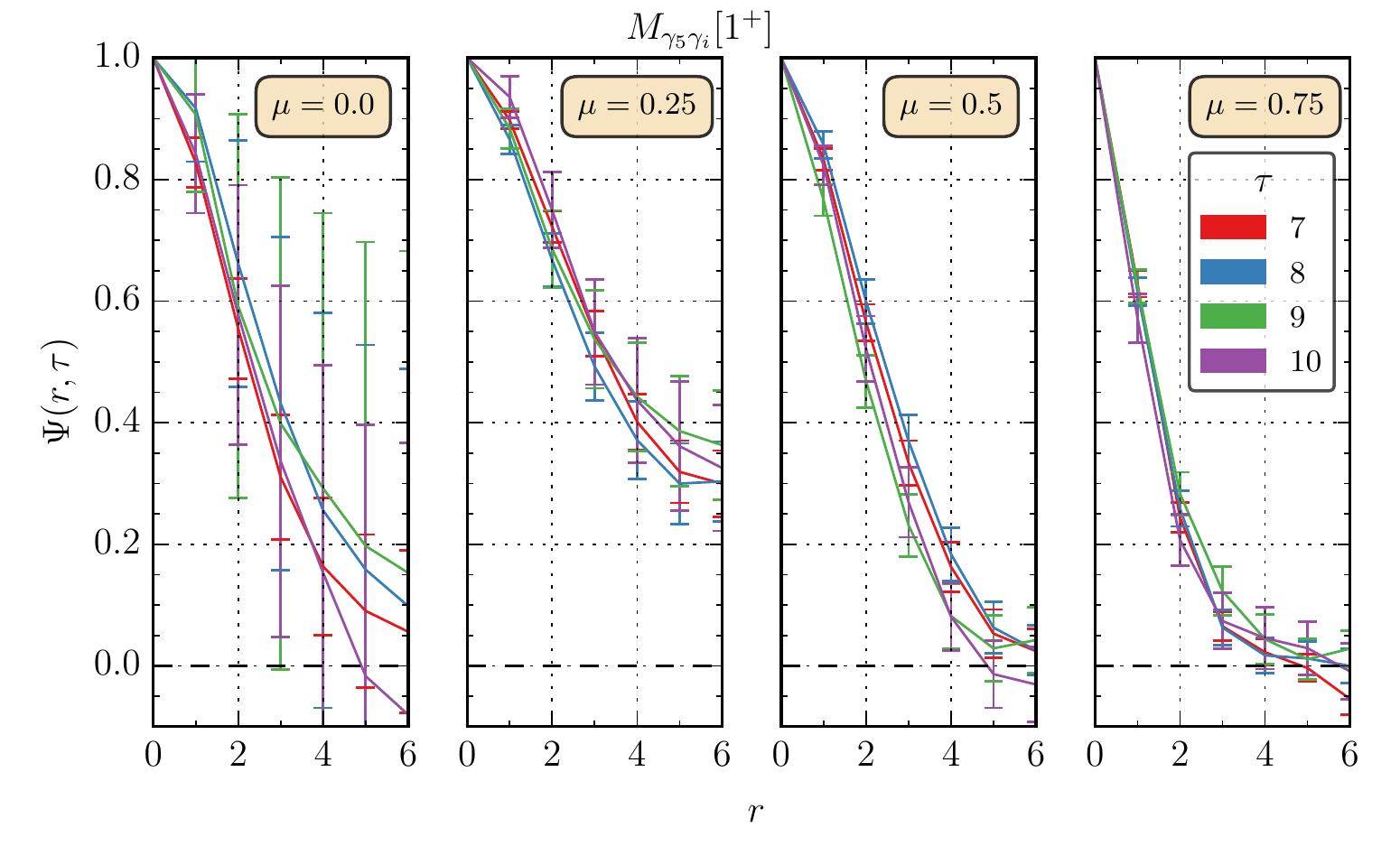}
\caption{Comparison between diquark (top) and meson (bottom) wavefunctions evaluated at different timeslices values $ \tau=7,8,9,10$. We show the results for $ 4 $ different values for the chemical potential $\mu=0.0,0.25,0.5,0.75$.}
\label{fig:wave_tau}
\end{center}
\end{figure}

Fig.~\ref{fig:wave_tau}  shows
wavefunctions calculated with varying $\tau$ values. 
For clarity's sake only
data from separations $\vec r$ on-axis are plotted. The $1^-$ diquark and the $1^+$ meson show similar behaviour; the approximate consistency of
data from different $\tau$ is good evidence for the existence of a discrete 
bound state and the applicability of the approximation (\ref{eq:wvfn}), although
for $\mu a\lesssim0.5$ the data is much nosier for the meson than the diquark.
For $\mu<\mu_o$ the wavefunction has a large width and has not yet vanished by
the lattice midpoint, which is consistent with the volume dependence of the
spin-1 data seen in Fig.~\ref{fig:voldep}. 

\begin{figure}[t]
\begin{center}
    \includegraphics{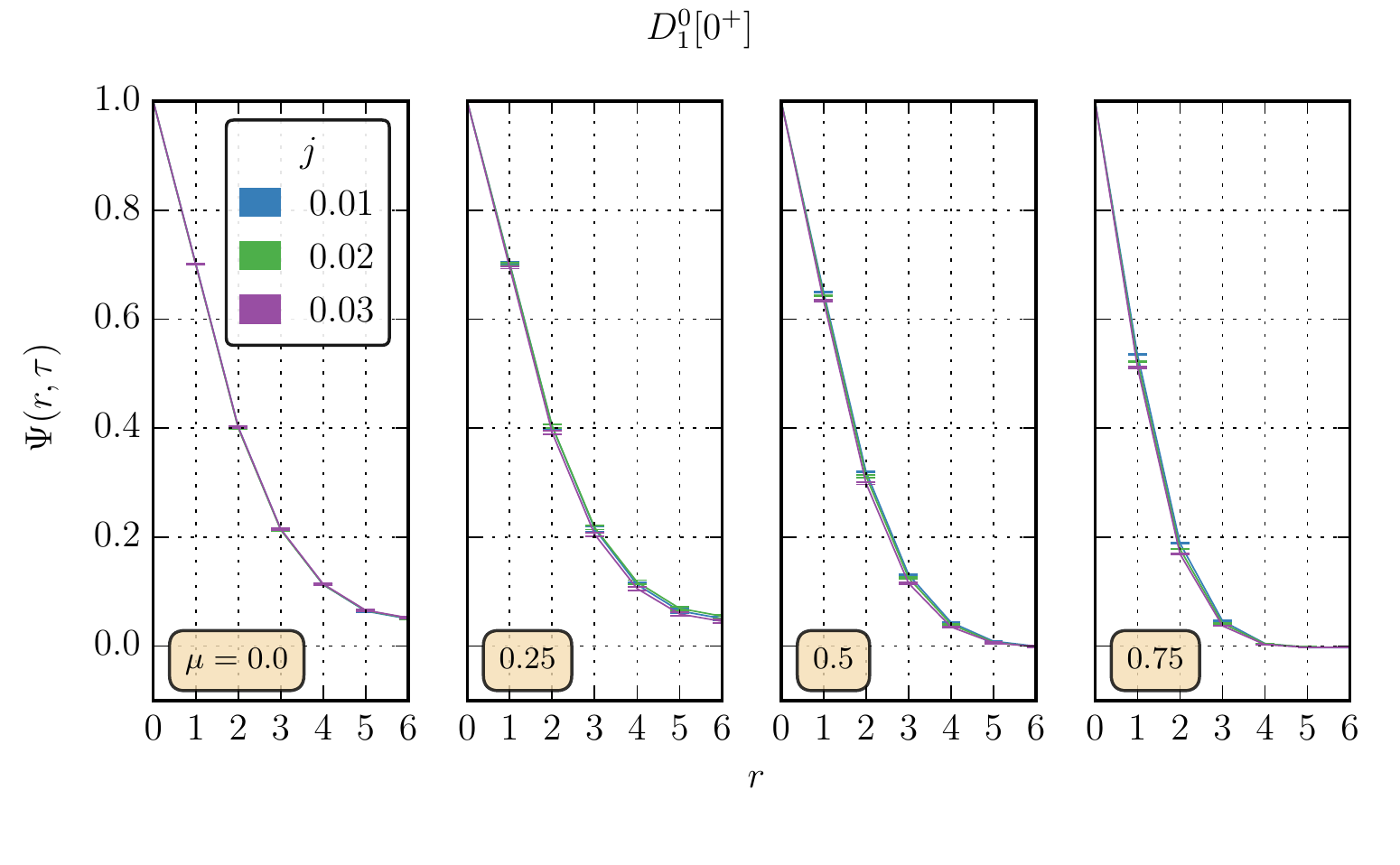}
\caption{Diquark wavefunctions relative to $ D^0_1 $ showing the $ j $-dependence for various values of the chemical potential $ \mu $.}
\label{fig:vary-j}
\end{center}
\end{figure}
In Fig.~\ref{fig:vary-j} we show the consequences for the scalar diquark
wavefunction of varying the diquark source
$j$; the data have been generated in a partially-quenched approach using
an ensemble  generated with $ja=0.02$. It can be seen that for $\mu>0$
increasing $j$ has the effect of very slightly shrinking the wavefunction. In
contrast to the data for free fermions in Fig~\ref{fig:free}, there is no sign
of oscillatory behaviour developing in the $j\to0$ limit.

\begin{figure}[p]
\begin{center}
    \includegraphics{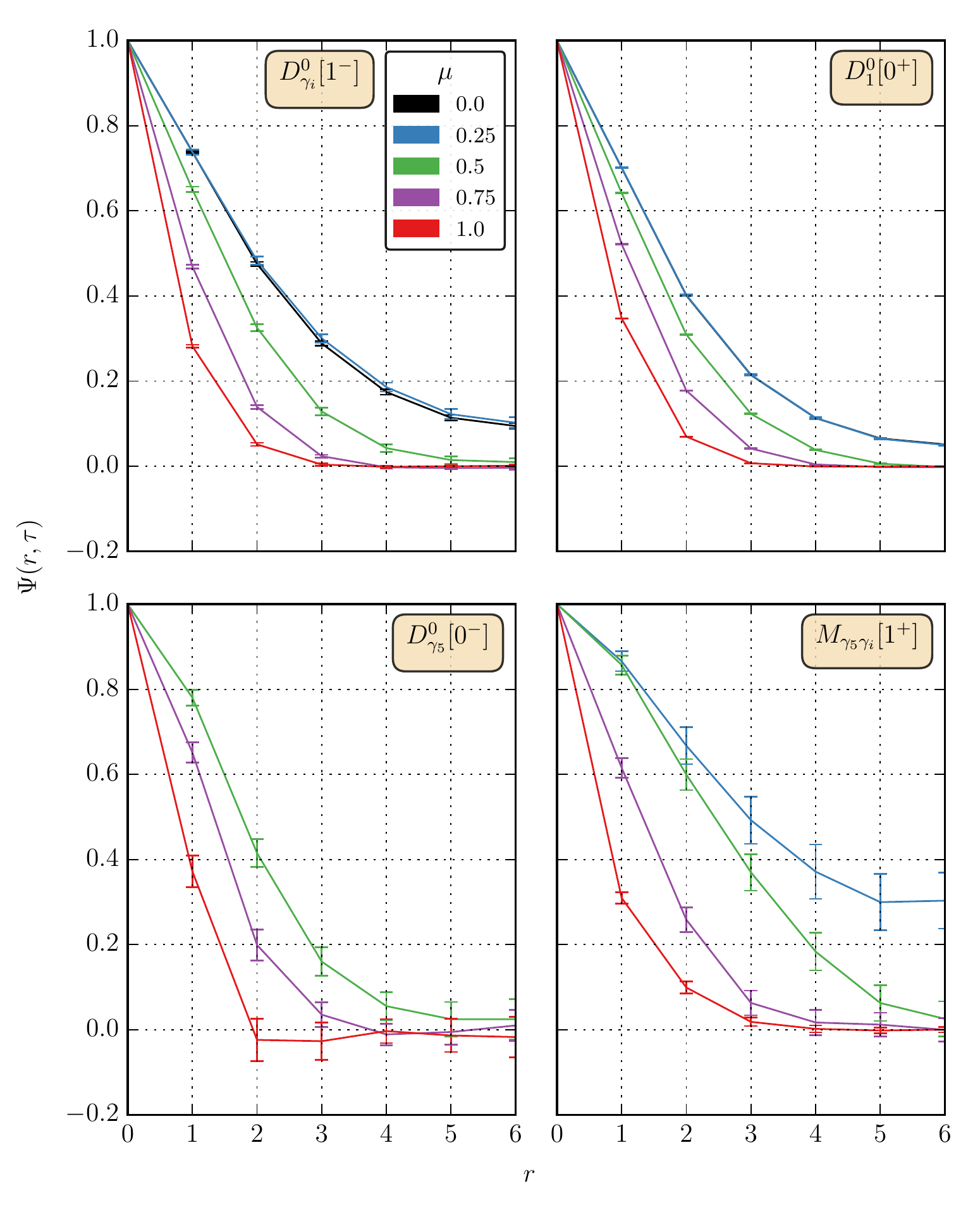}
\caption{Diquark and meson wavefunctions for various values of the chemical
potential $ \mu $ at $ja=0.02$. Below onset, some operators are very
noisy and are excluded from the plots for clarity.  }
\label{fig:waves}
\end{center}
\end{figure}
Figs.~\ref{fig:waves} show $\tau=8$ 
wavefunctions for the four channels of interest as $\mu$ is varied. Comparison
of the plots enables us to order the states by spatial size at $\mu=0$:
$0^+<1^-<1^+$. The $0^-$ data are
considerably noisier. The main common trend is the systematic decrease
in the size of all states as $\mu$ increases, so that by $\mu a\sim O(1)$ the
maximum extent $ra\sim3$. With the exception of the noisy
$0^-$, there is no sign of any of the wavefunctions changing sign or developing
oscillatory behaviour as $\mu$ increases. 

\begin{figure}[t]
 \includegraphics{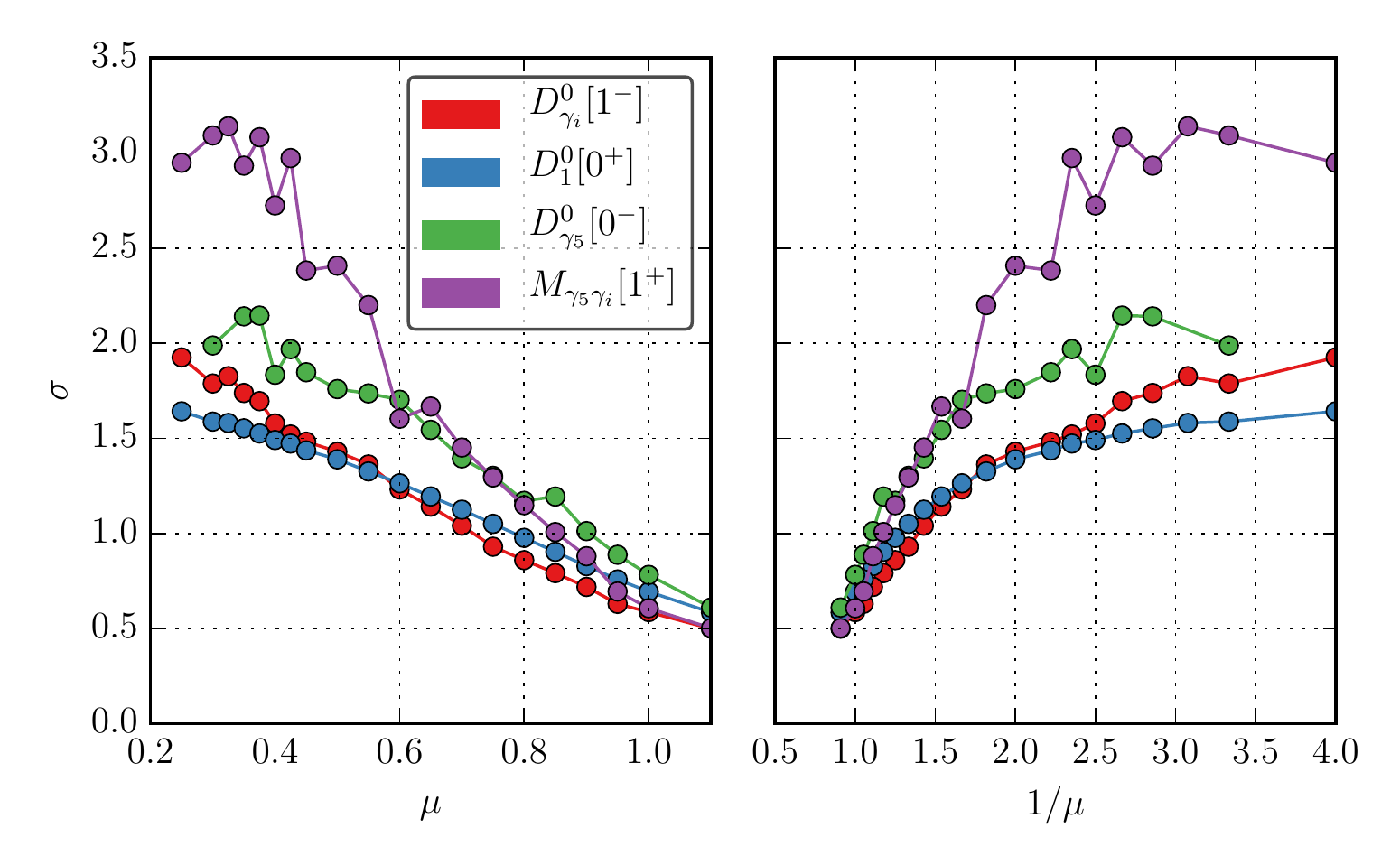} 
\caption{ Estimated width $\sigma$ of the wavefunction for
the different operators, obtained by fitting to a spline 
and then extracting the width at half maximum, as a function of $\mu$ (left) and
$\mu^{-1}$ (right).}
\label{fig:fit_width-s12t24}
\end{figure}

To crudely quantify the evolving spatial size, we extracted from the data the
full width at half the maximum of $\Psi(r,\tau)$, which was obtained by fitting
the wavefunction to a spline. In Fig.~\ref{fig:fit_width-s12t24} we then plot
the resulting width $\sigma(\mu)$ in each channel. The results are plotted both
as a function of $\mu$ and of $1/\mu$, and confirm the trends
reported above, and also highlights that post-onset $\sigma(0^-)$ is larger than
both $0^+$ and $1^+$ states. We note that $\sigma(0^+)>\sigma(1^-)$ for $\mu
a\gtrsim 0.6$, which is confirmed by close inspection of Fig.~\ref{fig:waves}.
For $\mu a\gtrsim 0.6$, or $1/\mu a\lesssim 1.5$, the widths approach one another making the
various channels difficult to distinguish, but the right panel of
Fig.~\ref{fig:fit_width-s12t24} suggests that bound states are
increasingly dominated by a single length scale $\sigma\propto\mu^{-1}$, with
the hierarchy $\sigma(1^-)<\sigma(0^+)<\sigma(1^+)<\sigma(0^-)$.

\section{Discussion}
\label{sec:discussion}
We have presented results from the first attempt, using orthodox lattice
techniques, to examine the spatial structure of
gauge-invariant excitations in a baryonic medium. The results complement a
previous study \cite{Hands:2007uc} of the excitation spectrum. We have focused
on the post-onset regime $\mu>\mu_o$ in which baryon charge density is non-zero
in the $T\to0$ limit, and the ground state is a superfluid. Our results are
consistent with
the indistinguishability of mesons and diquarks in a superfluid, and suggest a
scale hierarchy $\sigma(0^+)\sim\sigma(1^-)<\sigma(0^-)<\sigma(1^+)$, 
to be compared with the mass hierarchy
$m(0^+)<m(1^+)\ll m(1^-)<m(0^-)$ found in \cite{Hands:2007uc}.
As a general rule, signals obtained in diquark channels were less noisy than
those from mesons.

The scale hierarchy becomes less well-defined as $\mu$ increases and the
wavefunctions shrink; from $\mu
a\sim0.6$ onwards all the channels yield wavefunctions of approximately equal
extent. For orientation, if the string tension is used to set the scale this
corresponds to $\mu\simeq670$MeV, at which point the quark density $n_q\simeq
n_{SB}\sim5\mbox{fm}^{-3}$ (using (\ref{eq:SB})), or roughly 10$\times$ nuclear
density~\cite{Cotter:2012mb}. Spline fits to the profiles then yield the
approximate behaviour $\sigma\propto\mu^{-1}$ with a different hierarchy
 $\sigma(1^-)<\sigma(0^+)<\sigma(1^+)<\sigma(0^-)$.
This is consistent 
with the expectation $\sigma\propto\mu^{-1}$ which assumes
that $\mu\sim k_F$ is the only relevant scale at high density. 
Indeed, this is precisely the content of the free-field prediction
(\ref{eq:free_bessel}).
The physical picture is that bound-state excitations are formed from quarks
close to the Fermi surface with a characteristic de Broglie wavelength
$\lambda_F\sim\mu^{-1}$. The absence of appreciable finite volume effects
suggests, however, that the influence of image charges is negligible, and that
confinement continues to hold. The conjunction of both properties characterises the
so-called {\it quarkyonic\/} regime. 

Finally, the absence of oscillatory behaviour in $\Psi(r,\tau)$ at low temperature,
in contrast with the weak-coupling prediction (\ref{eq:free_bessel}) needs some
consideration. One obvious departure from weak-coupling behaviour is the
formation of a superfluid condensate $\langle qq\rangle\not=0$, which for a
degenerate system should, via the BCS mechanism, induce an energy gap
$\Delta\sim\Lambda_{\rm QC_2D}\sim\langle
qq\rangle/\mu^2$ at the Fermi surface. The presence of a gap removes the sharp
momentum cutoff in the integrals leading to the expressions
(\ref{eq:free_bessel},
\ref{eq:friedel_int}), and therefore also the oscillations. A gap can also
be modelled in free-field theory by the introduction of a diquark source
$j\not=0$, and  the curves shown in Fig.~\ref{fig:free} strongly suggest this
explicit gap does indeed dampen the oscillations. The fact that oscillations
remain absent from the wavefunctions of interacting quarks as $j\to0$, 
demonstrated in Fig.~\ref{fig:vary-j}, is consistent
with the post-onset QC$_2$D gap being generated dynamically.

\section{Acknowledgements}

This work is undertaken as part of the UKQCD collaboration using the STFC-funded
DiRAC Facility. The computational facilities of HPCWales were also used. 
The work of AA was supported by a postgraduate scholarship from Swansea University. AA also acknowledges European Union Grant Agreement number 238353 (ITN STRONGnet) and Academy of Finland grant 1267286 for support during the completion of this work.

\end{document}